\documentstyle{article}
\def\ov{\overline}

\begin{document}

\begin{flushright}
hep-ph/9907260\\
\end{flushright}

\begin{center}
{\Large \bf Neutral kaons as decay products and \\
analyzers of heavier flavors \footnote{Published in:
 {\it "Proceedings of the 14th International Workshop
 on High Energy Physics and Quantum Field Theory,
 Moscow, May 27--June 2, 1999", eds. B.Levchenko, V.Savrin,
 Moscow, 2000, pp.486-491}.  } }

\vspace{4mm}


Yakov I. Azimov\\

Petersburg Nuclear Physics Institute,\\
Gatchina, St.Petersburg, 188350, Russia \\

\end{center}


\begin{abstract}

Cascade decays of heavy flavor hadrons to states with neutral kaons
are discussed as an instrument for detailed studies of the heavy
hadron properties. For neutral flavored mesons the well-known kaon
oscillations provide a unique experimental possibility of relating
lighter/heavier eigenstate masses to (approximate) even/odd eigenstate
$CP$-parities and to longer/shorter eigenstate lifetimes. As a result,
they allow to eliminate sign ambiguities of $CP$-violating parameters
and, therefore, to check the Standard Model (or to find New Physics).
Specifically for charmed hadrons, both neutral/charged mesons and
baryons, the secondary kaon oscillations permit unambiguous separation
of Cabibbo-allowed and doubly-suppressed decay amplitudes, including
measurement of their relative phases. For neutral $D$-meson decays the
kaon oscillations can also discriminate, again unambiguously, between
effects of $D$-meson mixing and interference of suppressed/allowed
amplitudes. Another problem discussed is the influence of kaon
$CP$-violation on the amplitude structure and on phenomenology of
$CP$-violation in heavy hadron decays.

\end{abstract}


\section {Introduction}

$CP$-violation in heavy flavors is becoming an object of experimental
studies~[1-3]. However, many related theoretical problems are not quite
understood yet. For example, it was noted only recently (see, e.g.~[4-8])
that some discrete ambiguities may appear when extracting $CP$-violating
parameters from experiment and comparing them with theory. Such
ambiguities may shadow manifestations of New Physics (see discussion
in~\cite{gnw,w98}). Specific problems arise also in decays of $D$-mesons,
where doubly Cabibbo-suppressed transitions may imitate flavor mixing
effects.

Discussed in this talk is the unique role of neutral kaons produced in
decays of heavier flavors. Well-studied strangeness oscillations are
sensitive to the relative initial content of $K^0$ and $\ov K^0$.
Therefore, they may be used to analyze detailed properties of the
decays and heavier flavor hadrons themselves, just as, say, asymmetric
decays of hyperons are used to analyze the hyperon polarization and
properties of hyperon production. At this way one becomes able
to eliminate any ambiguities of $CP$-violating parameters. Decays to
neutral kaons are capable as well to separate "right" and "wrong"
strangeness transitions (Cabibbo-allowed and Cabibbo-suppressed
amplitudes) for charmed and beauty hadrons, both neutral and charged
mesons or baryons,
again unambiguously.  Also discussed are unfamiliar manifestations of
kaon $CP$-violation in heavy meson decays.

\section {Amplitude ambiguities and their nature}

Let us consider, in the standard manner, decays of neutral flavored
spinless mesons $M$ and $\ov M$. The most popular way to search for
$CP$-violation is to study decays
\begin{equation} M(\ov M)\to X_{CP}
\end{equation}
into a state of definite $CP$-parity. $CP$-violation in the decays may
be described by the parameter \begin{equation}
\lambda_X=\frac{q_M}{p_M}\,\frac{A_{\ov MX}}{A_{MX}}\,.
\end{equation}
It is rephasing invariant and is commonly considered as unambiguous.
However, there is an intrinsic sign ambiguity hidden in this
parameter.  Ambiguities described in the literature look differently in
different papers (e.g.,~\cite{ars,gkn,bk}), but all of them are really
related to just this sign ambiguity.

We can reveal the ambiguity by expressing $\lambda_X$ through decay
amplitudes $A^{(1)}_X$ and $A^{(2)}_X$ of the eigenstates
$M^{(1)}=p_MM+q_M\ov M,\;M^{(2)}=p_MM-q_M\ov M.$
In such a way we obtain
\begin{equation}
\lambda_X=\frac{A^{(1)}_X-A^{(2)}_X}{A^{(1)}_X+A^{(2)}_X}\,,
\;\;\;\;{\rm
Re}\lambda_X=\frac{|A^{(1)}_X|^2-|A^{(2)}_X|^2}
{|A^{(1)}_X+A^{(2)}_X|^2}\,,\;\;\;\;{\rm Im}\lambda_X=\frac{2\,{\rm
Im}(A^{(1)}_X\,A^{(2)\ast}_X)}{|A^{(1)}_X+A^{(2)}_X|^2}\,.
\end{equation}
This expression clearly shows that $\lambda_X$ changes its sign
under interchange of $M^{(1)}$ and $M^{(2)}$.
Thus, to fix the sign of $\lambda_X$ we need
to identify who is who in the set of the eigenstates. In other
words, definition (2) should be appended by some physical labeling
for the eigenstates.

Consider the situation in more detail. Decays (1) go along two
branches: $M(\ov M)\to{M^{(1)}}\to{X_{CP}} $ and
$\;M(\ov M)\to M^{(2)}\to X_{CP}.$ They produce separate
contributions to the decay amplitude which can interfere. Time
distribution of any decay (1) contains two kinds of terms linear
in $\lambda_X$, both unambiguously measurable.  Direct
contributions of the two branches
combine into the term proportional to $${\rm
Re}\lambda_X\cdot\sinh\frac{(\Gamma^{(1)}- \Gamma^{(2)})t}{2}\,,$$
while interference of the branches gives the term proportional to
$${\rm Im}\lambda_X\cdot\sin(m^{(1)}-m^{(2)})t\,.$$ Structure of these
terms allows to formulate the ambiguity problem more explicitly.

There are three possible ways of labeling the eigenstates:
\begin{itemize}
\item Lifetime labeling identifies the states as longer or shorter
lived. Then the sign of $\Delta\Gamma$ is fixed by definition, so
Re$\lambda_X$ is experimentally unambiguous. But the sign of $\Delta m$
is generally unknown, and the sign of Im$\lambda_X$ appears to be
ambiguous.
\item $CP$-parity labeling identifies the states (though may be
approximately) as $CP$-even or $CP$-odd. Here we define (see, e.g.,
ref.\cite{ars}) that $M^{(1)}$ has the same (approximate) $CP$-parity
as the final state $X_{CP}$ if it decays to $X_{CP}$ more intensely
than $M^{(2)}$. This means that $|A^{(1)}| > |A^{ (2)}|$, and so this
definition fixes the sign of Re$\lambda_X$ through eq.(3). The sign of
$\Delta\Gamma$ becomes measurable, but the signs of $\Delta m$ and
Im$\lambda_X$ stay unknown.
\item Mass labeling identifies the states as heavier or lighter.
Here the sign of $\Delta m$ is fixed, and Im$\lambda_X$ may be measured
unambiguously. But such convention does not fix the sign of $\Delta\Gamma$,
and thus Re$\lambda_X$ has the sign ambiguity (see, e.g., ref.\cite{bk}).
\end{itemize}
It is evident now that all experimental sign ambiguities would be
eliminated if we could relate those three labelings to each other.

The lifetime and $CP$-parity labelings can be related to each other
in a straightforward way by comparing time-dependences of decays into
final states of different $CP$-parities. However, their relation to
the mass labeling is not so simple.

We can illustrate the general situation by comparing it to the
well-studied case of neutral kaons.  Kaon eigenstates are defined at
present as $K_S$ amd $K_L$ through their lifetimes. Correspondence of
the lifetimes and $CP$-parities has been achieved in decays to $2\pi$
and/or $3\pi$.  Note that similar attempt was recently made also for
$D$-mesons~\cite{791}, but the achieved precision appeared still
insufficient to notice any difference of the two lifetimes.

Kaon mass labeling, i.e. identification of $K_L$ as the heavier
state, became possible only after special complicated experiments
on coherent regeneration (their summary see
in~\cite{PDTo}) which related to each other the masses and
$CP$-parities of kaon eigenstates.  Without such mass labeling
the standard $CP$-violating kaon parameters $\eta$ could be
measured only up to the sign of~$\,{\rm Im}\eta$.

For $B$- and $D$-mesons the coherent regeneration cannot be observed
because of their too short lifetimes.  Instead one may use some
theoretical assumptions (e.g.,~\cite{gkn}).  However, there should
exist direct experimental ways to relate all three kinds of labeling,
independently of any theoretical assumptions. One of them is described
in the next section. Interestingly enough and similar to how it was for
kaons, the direct experimental interrelation of eigenmasses and
eigenwidths appears to be impossible for $B$- and $D$-mesons as well
(the corresponding discussion see in~\cite{azD}). Both masses and
widths can be directly related only to $CP$-parities of the eigenstates,
and only after that to each other.

\section {Neutral kaons as analyzers of heavier flavors}

It is well known that weak decays of hyperons, being asymmetric
due to parity violation, are good analyzers which may be used
to measure hyperon polarization in various processes.  In
direct analogy, it was suggested~\cite{az1} that the neutral
kaons, with their decay oscillations, may be used to analyze
properties of heavy flavor hadrons. Here we only briefly explain
the main ideas underlying such an approach. More technical
details, with accurate formulas, may be found in~\cite{az1,azD}.

Instead of decays (1) we consider now decays
\begin{equation} M(\ov M)\to X_{CP}K^0(\ov K^0)
\end{equation}
and assume that $X_{CP}$ has definite values of both $CP$-parity and
spin. Generally, there are two possible kinds of flavor transitions,
$M\to K^0$ and $M\to\ov K^0$ (and two charge conjugate ones), with
corresponding different amplitudes. The very essential point is that
a definite coherent mixture of $\,M$ and $\ov M$ just before the decay
(4) produces some different, but also definite coherent mixture of
$\,K^0$ and $\ov K^0$ just after the decay. As a result, time evolution
of neutral kaons after decay (4) coherently continues evolution of
the flavored neutral mesons before the decay~\cite{aJL}.

The neutral kaons may be observed only through their decay, so we really
have cascade decays with two stages and two decay times, $t_M$ and
$t_K$ (we mean not average lifetimes, but event-by-event times).
Flavor oscillations at the two stages are correlated\footnote{ The term
"cascade mixing"~\cite{ks} looks not adequate, since mixings at the two
stages have the usual standard form and are not related to each other;
related are only flavor oscillations.}.  Such coherent double-flavor
oscillations produce generally non-factorisable dependence on the two
decay times.  It is rather complicated (see~\cite{az1,azD}), and for
better understanding its main features we may first simplify it by
assuming exact $CP$-conservation (for both $M$-mesons and kaons!). Then
the states $X_{CP}K_S$ and $X_{CP}K_L$ have definite $CP$-parities.
Eigenstates $M^{(1)}$ and $M^{(2)}$ also have definite $CP$-parities,
which may be used as their labels (just as $K_1$ and $K_2$ for
neural kaons before discovery of $CP$-violation). Of course, only two
transitions, say, \begin{equation} M^{(1)}\to X_{CP}K_L\,,\;\;\;
M^{(2)}\to X_{CP}K_S \end{equation} are possible here, instead of four
ones in the general case (initial and final $CP$-parities should be the
same).

If we observe the secondary kaons by their decays to $2\pi$ or $3\pi$
modes, then only one of transitions (5) contributes (recall the
assumption of exact $CP$-conservation!). As a result, dependences on
$t_M$ and $t_K$ become factorised. Such decay chains allow to relate
lifetimes and $CP$-parities of $M^{(1,2)}$. But if we use semileptonic
kaon decays, with contributions from both $K_L$ and $K_S$, then both
transitions work; time distribution contains their interference term
proportional to $$\cos(\Delta m_Mt_M-\Delta m_Kt_K)\,.$$ Its
measurement, evidently, determines the sign of $\Delta m_M$ in respect
to the known sign of $\Delta m_K$ and, therefore, relates masses and
$CP$-parities of $M$-eigenstates.

$CP$-violation complicates this picture due to interferences of all four
cascade branches~\cite{az1,azD}. Nevertheless, studies of decays (4)
are still capable to relate mass labeling of heavy meson eigenstates
with their $CP$-parity labeling and, then, with lifetime one.
Therefore, as explained above, such studies can eliminate all
experimental ambiguities in $CP$-violating parameters.

Decays of charmed particles suggest one more test-ground for the
analyzing power of neutral kaon oscillations. As a rule, $B$-meson
decays to neutral kaons have only one flavor transition (and its charge
conjugate), $B\to K$ or $B\to\ov K$ (compare decays considered in
refs.~\cite{az1} and~\cite{ad}). On the opposite, $D$-mesons, neutral
and charged (and even charmed baryons), always have both kinds of
transitions. One of then is doubly Cabibbo-suppressed, i.e. relatively
small (about 3\%  in the amplitude). Nevertheless, it is of special
interest: it exemplifies a new kind of weak transitions and might
demonstrate different (larger?) $CP$-violation.  The same is true, of
course, for decays to charged kaons as well, which may produce kaon
of$~$ "wrong" sign.  However, decays to neutral $K$'s have an essential
difference.  Final states for the charged kaon case are not coherent,
and one can compare only absolute values of the decay amplitudes.  As
was first noticed in~\cite{aj} (see also ref.\cite{s1}), decays to
neutral kaons produce coherent states and allow to measure as well the
relative phase of the amplitudes.

The current literature contains suggestions to realize this by measuring
probabilities of transitions $D\to K_S\,,\,\,D\to K_L$~\cite{by,xing}.
But in such an approach the relative phase of the transition amplitudes
cannot be measured, and separation of the amplitudes for $D\to\ov K^0,
\,D\to K^0$ appears to be ambiguous. Measurement of strangeness
oscillations for the secondary neutral kaons makes the amplitude
separation for different flavor transitions quite unambiguous (more
details see in~\cite{azD}; similar ideas were discussed in~\cite{s1,s2}).
We emphasize that the oscillations may provide as well clear separation
of two sources of$~$ "wrong" strangeness production, the
Cabibbo-suppressed transition and mixing of initial neutral $D$-mesons.

\section{Secondary kaon $CP$-violation}

Manifestations of$~$  kaon $CP$-violation in kaon decays has
been studied in all detail, at least phenomenologically.
However, production of neutral kaons (in particular, in decays
of heavier particles) provides different manifestations, not
quite familiar. Some of them  look rather formal at present, b
ut may become physically meaningful in future experiments.
Moreover, they might be useful for studying $CP$-violation
related to heavier hadrons. Here we consider two kinds of
such manifestations: for amplitudes, and for decay yields.

We begin with the problem, what are amplitudes for production of
$\,K_{S,L}$, say, in decays (4) of the meson $M$. Since $K_{S,L}=p_K
K^0\pm q_K\ov K^0$, it seems natural to express those amplitudes
through the amplitudes $A^{(X)}_{MK}$ and $A^{(X)}_{M\ov K}$ of
flavor transitions $M\to K^0$ and $M\to\ov K^0$ as
\begin{equation} p^\ast_K A^{(X)}_{MK}\pm q^\ast_K A^{(X)}_{M\ov
K}\,.  \end{equation}
Such expressions indeed exist in the literature, but they are
incorrect. To understand why and to find correct expressions we
first recall the general meaning of amplitudes.

When given an initial state $|i>$ and the $S$-matrix, the amplitudes
$A_{ik}$ of transitions $|i>\,\to|k>$ are defined by decomposition of
the final state $S|i>$ in terms of some set of states $|k>$:
$$S|i>\,=\sum_k A_{ik}|k>\,.$$ If the set is orthonormalized we arrive
at the canonical expression $A_{ik}=\,<k|S|i>$. However, this
expression is inapplicable if the states $|k>$ are not orthogonal.

Let us apply this consideration to decays (4). Decay of the meson $M$
produces the kaon state (up to normalization)
$$A^{(X)}_{MK}K^0+A^{(X)}_{M\ov K}\ov K^0\,.$$ To find amplitudes of
transitions $M\to K_{S,L}$ we should decompose this final state in terms
of $K_{S,L}$ and extract the corresponding coefficients. Since
$$K^0=(K_S+K_L)/(2p_K)\,,\;\;\;\ov K^0=(K_S - K_L)/(2q_K)\,,$$ we
finally obtain \begin{equation}
A^{(X)}_{MS}=\frac{A^{(X)}_{MK}}{2p_K}+\frac{A^{(X)}_{M\ov K}}{2q_K}\,,
\;\;\;\;\;
A^{(X)}_{ML}=\frac{A^{(X)}_{MK}}{2p_K}-\frac{A^{(X)}_{M\ov K}}{2q_K}\,.
\end{equation}
Amplitudes for decays of $\,\ov M$ may be found in the same way.

Expressions (6) and (7) coincide if $|p|^2=|q|^2=1/2$, i.e. when $CP$
is conserved and the states $K_{S,L}$ are orthogonal. They differ
when $CP$ is violated and the states $K_{S,L}$ are not orthogonal.
Using decomposition over non-orthogonal set of states is non-standard,
but looks natural for kaons with $CP$-violation. Expressions (7) just
correspond to those used earlier in refs.~\cite{aj,az1,azD}. Of course,
expressions (6) and (7) for amplitudes lead to different expressions
for decay probabilities as well.

There is one more consequence of kaon $CP$-violation for processes
with neutral kaons produced. Let us compare decays of initially pure
states $K^0$ and/or $\ov K^0$ into a particular mode. Their time
dependences generally oscillate. The oscillations would be absent for
some decay modes if $CP$ were conserved; they are present for any mode
when $CP$ is violated. These oscillations are different for initial
$K^0$ or $\ov K^0$. As a result, decay yields at a particular time
moment (and even total decay yields) are also different for initially
pure $K^0$ or $\ov K^0$. Of course, the relative difference is of the
order $|\eta|$. Similar difference, generally, exists for any coherent
mixture of $K^0$ and $\ov K^0$ having no symmetry under their
interchange.

Now, compare decays of, say, $D$ (neutral or charged) and $\ov D$,
with neutral kaons produced. Assume that the $D$-decay generates a
kaon system $aK^0+b\ov K^0$. The conjugate $\ov D$-decay without any
$CP$-violation, direct or in $D$-mixing, generates the conjugate system
$a\ov K^0+bK^0$.  Using any particular way to detect the secondary
neutral kaons (e.g., particular decay modes and/or particular
interval(s) of kaon decay times) will lead to different results for
$D$ and $\ov D$ due to kaon $CP$-violation. The difference is still
present after integration over $t_K$.  Theoretically, such
$t_K$-integrated effect was first demonstrated some years ago~\cite{aj}
for the sequence $D^\pm\to\pi^\pm K^0(\ov K^0),\;K^0(\ov K^0)\to\pi^+
\pi^-$ . Recent (but not quite correct) discussion of the small-$t_K$
region see in~\cite{lx}. We emphasize that similar effects of the kaon
$CP$-violation should appear in all decays of heavier flavor hadrons,
both mesons and baryons, to neutral kaons.

Presence of $D$-meson $CP$-violation does not eliminate the discussed
effect.  Moreover, kaon $CP$-violation in cascade decays appears to
be coherent with the $D$-meson $CP$-violation and may be used to
analyze its details.  Thus, we may have one more example of the
analyzing power of neutral kaons.

\section{Conclusion}

In summary, we see that neutral kaons being decay products may provide
the great analyzing power for very detailed studies of heavier flavor
hadrons and their decays.

\section*{Acknowledgments}

Correspondence with B.Kayser, H.Lipkin, Y.Nir, J.P.Silva and Z.-Z.Xing
was useful for me in preparing this talk.



\begin{thebibliography}{99}

 \bibitem{CPD}  E.M. Aitala et al., Phys.Lett. 421B (1998) 405;
e-print hep-ex/9711003.
\vspace{-2.5mm}
 \bibitem{Hit}  H. Yamamoto, e-print hep-ph/9812279.\\
        M. Artuso, e-print hep-ph/9906379.
\vspace{-2.5mm}
 \bibitem{OPAL} K. Ackerstaff et al., Eur.Phys.J. C5 (1998) 379;
e-print hep-ex/9801022.\\ M.P. Schmidt, e-print hep-ex/9906029.
\vspace{-2.5mm}
 \bibitem{ars}  Ya.I. Azimov, V.L. Rappoport, V.V. Sarantsev,
         Z.Physik A356 (1997) 437;\\
e-print hep-ph/9608478.
\vspace{-2.5mm}
 \bibitem{gnw}  Y. Grossman, Y. Nir, M. Worah, Phys.Lett.
407B (1997) 307; e-print hep-ph/9704287.
\vspace{-2.5mm}
 \bibitem{gq}   Y. Grossman, H.R. Quinn, Phys.Rev. D56 (1997) 7259;
e-print hep-ph/9705356.
\vspace{-2.5mm}
 \bibitem{gkn}  Y. Grossman, B. Kayser, Y. Nir, Phys.Lett.
415B (1998) 90; e-print hep-ph/9708398.
\vspace{-2.5mm}
 \bibitem{bk} B. Kayser, talk at the Moriond Workshop on Electroweak
               Interactions and Unified Theories,\\ Les Arcs, France,
               March 1997; e-print hep-ph/9709382.
\vspace{-2.5mm}
 \bibitem{w98}  L. Wolfenstein, Phys.Rev. D57 (1998) 6857;
                e-print hep-ph/9801386.
\vspace{-2.5mm}
 \bibitem{791}  E791 Collaboration, Preprint FERMILAB-Pub-99/036-E,
  1999; e-print hep-ex/9903012.
\vspace{-2.5mm}
 \bibitem{PDTo} Particle Data Group, Phys.Lett. 170B (1986) 132.
\vspace{-2.5mm}
 \bibitem{az1}  Ya.I. Azimov, Phys.Rev. D42 (1990) 3705.
\vspace{-2.5mm}
 \bibitem{azD}  Ya.I. Azimov, Eur.Phys.J. A4 (1999) 21;
                e-print hep-ph/9808386.
\vspace{-2.5mm}
 \bibitem{aJL}  Ya.I.Azimov, Pis'ma ZhETF 50 (1989) 413
                [JETP Lett. 50 (1989) 447].
\vspace{-2.5mm}
 \bibitem{ks}   B. Kayser, L. Stodolsky, e-print hep-ph/9610522.
\vspace{-2.5mm}
 \bibitem{ad}   Ya.I. Azimov, I. Dunietz, Phys.Lett. 395B(1997) 334;
e-print hep-ph/9612265.
\vspace{-2.5mm}
 \bibitem{aj}   Ya.I. Azimov, A.A. Johansen, Yad.Fiz. 33 (1981) 388
 [Sov.J.Nucl.Phys. 33 (1981) 205].
\vspace{-2.5mm}
 \bibitem{s1}   C.C. Meca, J.P. Silva, Phys.Rev.Lett. 81 (1998) 1377 ;
   e-print hep-ph/9807320.
\vspace{-2.5mm}
 \bibitem{by}   I.I. Bigi, H. Yamamoto, Phys.Lett. 349B (1995) 363 ;
   e-print hep-ph/9502238.
\vspace{-2.5mm}
 \bibitem{xing} Z.-Z. Xing, talk at the 2nd Intern.Conf. on $B$-physics
               and $CP$-violation, Honolulu, Hawaii,\\ March 24-27,
1997; e-print hep-ph/9703459.
\vspace{-2.5mm}
 \bibitem{s2}   A. Amorim, M.G. Santos, J.P. Silva, Phys.Rev.
  D59 (1999) 056001; e-print hep-ph/9807364.
\vspace{-2.5mm}
 \bibitem{lx}   H. Lipkin, Z.-Z. Xing,  e-print hep-ph/9901329.




\end{thebibliography}
\end{document}